\documentclass[aps,prd,reprint,amsmath,amssymb]{revtex4-2}
\usepackage{inputenc}
\usepackage[english]{babel}
\usepackage{graphicx,color}
\usepackage{wrapfig}
\usepackage{microtype}
\usepackage{enumitem}
\usepackage{dsfont}
\usepackage{siunitx}
\usepackage[caption=false]{subfig}
\usepackage{textcomp}
\usepackage{stmaryrd}
\usepackage{float}
\usepackage{hyperref}
\usepackage{titlesec}
\usepackage{comment}
\usepackage{gensymb}
\usepackage{braket}
\usepackage{mathtools} 
\usepackage{cancel}

\DeclarePairedDelimiter\autobracket{(}{)}
\newcommand{\br}[1]{\autobracket*{#1}}

\DeclarePairedDelimiter\autobrackett{[}{]}
\newcommand{\brr}[1]{\autobrackett*{#1}}

\def\vp{{\boldsymbol{p}}}

\def\vA{{\boldsymbol{A}}}
\def\vE{{\boldsymbol{E}}}

\def\tr{{\text{r}}}
\def\ta{{\text{a}}}
\def\tk{{\text{k}}}

\newcommand\varpm{\mathbin{\vcenter{\hbox{\oalign{\hfil$\scriptstyle+$\hfil\cr\noalign{\kern-.3ex}$\scriptscriptstyle({-})$\cr}}}}}

\hypersetup{pdfencoding=auto}

\DeclareMathOperator{\Tr}{Tr}
\DeclareMathOperator{\sign}{sign}

\begin{document}

\title{Anderson-Higgs amplitude mode in Josephson junctions}

\author{P. Vallet}
\email{pierre.vallet@u-bordeaux.fr}
\affiliation{Univ. Bordeaux, CNRS, LOMA, UMR 5798, F-33400, Talence, France}

\author{J. Cayssol}
\email{jerome.cayssol@u-bordeaux.fr}
\affiliation{Univ. Bordeaux, CNRS, LOMA, UMR 5798, F-33400, Talence, France}

\date{\today}

\begin{abstract}
The Anderson-Higgs mode in a superconductor corresponds to a collective and coherent oscillation of the order parameter amplitude. We propose to detect this mode in a tunnel Josephson junction between two singlet s-wave diffusive superconductors. We find a strong enhancement of the tunneling current when the junction is pumped at the equilibrium gap frequency, corresponding to the activation of the Anderson-Higgs mode. By solving the Keldysh-Usadel equations, we obtain current peaks at specific bias voltages that can serve as signatures of the Anderson-Higgs mode.\end{abstract}

\maketitle

\section{Introduction}
 
\begin{figure}
    \centering
    \includegraphics[width=\columnwidth]{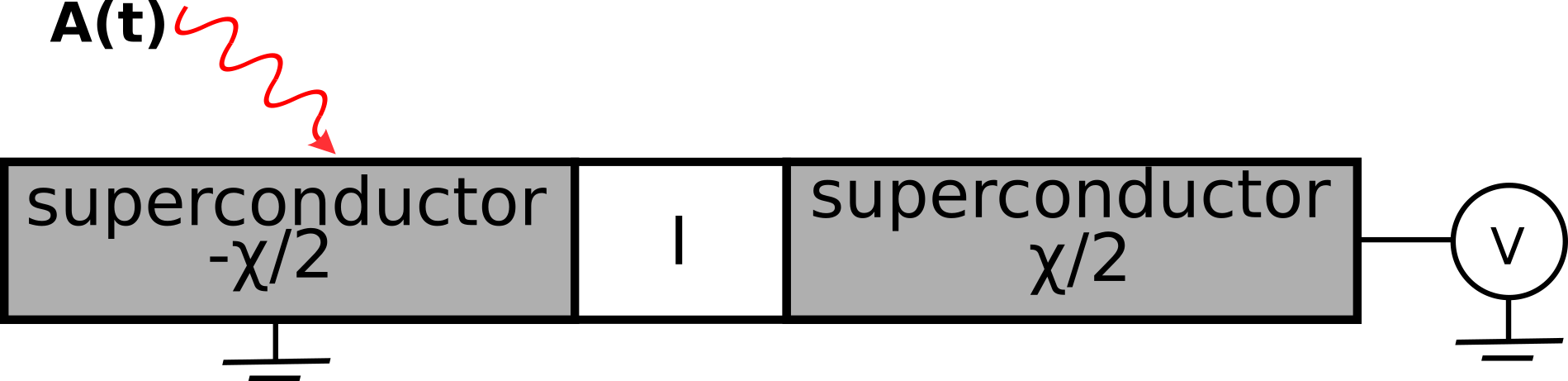}
    \caption{Josephson junction between two diffusive superconducting leads with a phase difference $\chi=\theta_R - \theta_L$. The left SC is irradiated by a monochromatic THz light to excite the Higgs mode. The right SC is put to a dc bias V.}
    \label{fig:sis}
\end{figure}

The Josephson effect is a hallmark of superfluidity and superconductivity. Both superfluids and superconductors are characterized by a complex valued order parameter $\Delta e^{i\theta}$ which can be interpreted as a macroscopic wave function \cite{Tinkham1996}. The absolute phase $\theta$ is non-measurable by itself, but a phase gradient leads to an observable supercurrent flow. In particular, when two superconductors, with respective phases $\theta_L$ and $\theta_R $, are separated by a weak-link, an equilibrium current $I(\chi)$ flows without dissipation from one lead to the other \cite{Josephson1962Jul}. In such a Josephson junction, the supercurrent is a flow of Cooper pairs which is controlled by the phase difference $\chi=\theta_R - \theta_L$. This Josephson effect has a longstanding and on-going history yielding to a variety of fundamental effects and applications in nanotechnology, including sensors and qubits. The weak-link can be a thin insulating layer, an atomic contact \cite{Bretheau2017}, a carbon nanotube \cite{Kazumov1999,Pillet2010}, a graphene single layer \cite{Heersche2007,BenShalom2016,Bretheau2017} or even a topological insulator \cite{Bocquillon2017Feb}, thereby allowing to probe the quantum properties of the link via the Josephson effect. The original and simplest version of a Josephson junction is the SIS junction where the two superconducting leads are separated by a thin insulating layer (Fig. \ref{fig:sis}). By applying a voltage difference $V$ between the 2 superconductors, the phase difference precesses at angular frequency $ \omega_J =  2eV/\hbar$. At finite bias, a transparent Josephson junction exhibits ac Josephson effect with subgap harmonic structures consisting of peaks located at $eV = 2\Delta /n$, in the differential conductance $dI/dV$ \cite{Arnold1987Jul, Kleinsasser1994Mar, Averin1995Aug}, $n$ being an integer number. Those structures originate from Multiple Andreev Reflection (MAR) and does not need any irradiation. If furthermore the junction is irradiated at frequency $\omega$, additional satellite peaks appear at biases $eV = 2\Delta /n \pm m \hbar \omega /n$ \cite{Gregers-Hansen1973Aug,Hasselberg1974May, Hansen1979Jul}, $m$ being another integer number associated to photon assisted MAR \cite{Cuevas2002Mar}. In a tunnel junction only the simple Andreev Reflection (AR) corresponding to $n=1$ has visible effect on the differential conductance. In this case, irradiated SIS junctions only exhibit PAT structures at biases $eV = 2\Delta \pm m \hbar \omega$. When a junction is simultaneously irradiated and biased, a resonance due to synchronization of the SCs phase change with the external excitation happens when the Josephson frequency $\omega_J$ is a multiple of the drive frequency $\omega$, resulting in Shapiro peaks in the dc current \cite{Shapiro1964Jan}. All these effects have mostly been studied within the microwave electromagnetic range $\omega \sim \text{GHz}$.

Another fundamental phenomenon in physics is the mechanism of spontaneous symmetry breaking of a local $U(1)$ gauge symmetry  \cite{Weinberg1996Sep}, uncovered by Anderson \cite{Anderson1963Apr} in superconductors and adapted to more general field theories and to the standard model \cite{Higgs1964Oct,Littlewood1982Nov}. A neutral superfluid has both Goldstone modes associated to soft global rotations of phase $\theta$ and an amplitude mode corresponding to  coherent oscillation of the superconducting gap amplitude $\Delta(t)$. Anderson discovered that for a charged superfluid, like a BCS superconductor, the Goldstone mode is pushed to higher energies, while the photon acquires a mass, which is the Meissner effect. Then the lowest-energy collective mode is the amplitude mode with mass $\hbar \omega_H = 2 \Delta_0$ in the meV range, corresponding to the gap of conventional superconductors. The discovery of the Higgs mode in condensed matter was realized due to the development of intense THz laser sources \cite{Hebling2002Oct, Hebling2008Jul, Hirori2011Feb, Fulop2012Feb}. Setting aside indirect measurements in system with charge density wave \cite{higgs_1980} using Raman spectroscopy, several pump-probe experiments have reported a Higgs signal via third harmonic generation \cite{matsunaga_2013_NbN,matsunaga_2014_NbN,Higgs_dhightc,Chu2020}. Besides those all-optical experiments, Tang et al. \cite{Tang2020Jun} proposed recently an electronic transport protocol based on the differential conductance of a normal metal/insulator/superconductor interface (NIS). Other signatures of the Higgs mode has been predicted in hybrid systems \cite{Silaev2020,Vallet2023Sep} or in irradiated bulk superconductors \cite{Plastovets2023,Derendorf2024}. 

The interplay of Josephson effect and Anderson-Higgs mechanism has not been considered yet, except in a recent preprint \cite{Lahiri2024Feb} predicting that the Higgs mode can be excited internally, in the absence of any irradiation, by the Josephson precession at $\omega_J$ of the phase in a transparent Josephson junction. In contrast, the interplay of i) THz external pumping, ii) Josephson precessing and iii) Higgs mode oscillation has not been considered, even in the simplest version of a Josephson element, namely in a SIS tunnel junction. Besides, the Higgs response is stronger in diffusive superconductors compared to clean ones \cite{Tang2020Jun,Vallet2023Sep}.

In this work we investigate Higgs mode signatures in a diffusive SIS Josephson junction pumped by an electromagnetic drive in the THz range. This external drive resonantly activates the Higgs mode via nonlinear coupling. We find a strong enhancement of the Josephson current for any bias voltage at the frequency drive corresponding to the equilibrium gap of the superconductors. Moreover the current-voltage characteristics $I(V)$ are predicted to exhibit a distinctive pattern of peaks caused by photon-assisted transmission (PAT) and a Shapiro peak at bias $eV = \hbar \omega$.  

The paper is organized as follows. In Sec. \ref{sec:model}, the Josephson junction model and the formalism is presented.  
Then the Keldysh-Usadel equations are solved for the non-irradiated SIS tunnel junction in Sec.\ref{sec:non_irradiated}, thereby allowing to write down the zero-order building blocks for the perturbative calculations in the irradiated case. In Sec. \ref{sec:irradiated}, we present our results on the Josephson tunnel current through the irradiated junction.   

\section{Model}\label{sec:model}

In this section, we present the tunnel Josephson junction model and the formalism used to study the Josephson effects in presence of a Higgs mode generated by the irradiation.

\subsection{SIS junction and parameters}

We consider a SIS Josephson junction consisting in two conventional s-wave singlet SCs separated by an insulating tunnel barrier (Fig. \ref{fig:sis}). Our system can be viewed as the Josephson version of the NIS study by Tang et al. \cite{Tang2020Jun}. Both superconductors are in the diffusive regime, and a dc voltage bias is also applied through the junction. To excite the Higgs mode, one of the superconducting lead is pumped by a THz electric field $\vE = -\partial_t \vA$ where $\vA(t) = \vA_0 \br{e^{i\omega t} + e^{-i\omega t}}$ is a uniform harmonic vector potential with amplitude $\vA_0$. In a dirty diffusive SCs, a uniform potential vector with amplitude $|\vA_0|$ introduces a typical energy scale  
\begin{equation}
    A_D = \frac{D e^2 |\vA_0|^2}{\hbar} \, .
\end{equation}
 In the following, the irradiation strength is charaterized by the dimensionless ratio of energy scales $\alpha=A_D/\Delta_0$. Besides the electromagnetic drive at angular frequency $\omega$, the dynamics of the system is also characterized by the Josephson frequency $\omega_J = 2eV/\hbar$.
 
 So far, irradiated Josephson junctions have mostly been extensively studied in the regime $\hbar \omega \ll 2\Delta_0$, corresponding to microwave frequencies $\sim \text{GHz}$ \cite{Dayem1962Mar, Tien1963Jan, Shapiro1964Jan, Langenberg1966Apr, Gregers-Hansen1973Aug, Hansen1979Jul, Roychowdhury2015Sep, Gonzalez2020Jul, Carrad2022Aug,  Haxell2023Oct}. In this case no quasiparticles (QPs) nor Higgs mode can be excited by the electromagnetic drive. Irradiation produces Shapiro spikes at dc bias $V=n \hbar\omega /2e$ and photon assisted tunneling (PAT) \cite{Shapiro1964Jan, Roychowdhury2015Sep}. It also induces a reconstruction of the SC ground state \cite{Semenov2016Jul}. 

Here we consider higher frequency range allowing to excite the Higgs mode, typically $\hbar \omega \sim \Delta_0 $ which falls into the THz range \cite{Tsuji2015Aug}. Moreover, the calculations will be performed in a perturbative way with respect to the irradiation strength $\alpha$, namely in the regime 
\begin{equation}
    A_D \ll \hbar \omega \sim \Delta_0.
\end{equation} In this regime, the photon energies $\hbar \omega$ are still below the $2\Delta_0$ threshold mandatory to break Cooper pair. Nevertheless those photons can excite Higgs mode via a second order nonlinear process (Sec. \ref{sec:usadel}).

The low transparency of the tunnel layer between the two diffusive SCs is characterized by its normal state tunnel conductance. We we use circuit theory \cite{Nazarov1999May} to extract the Josephson current 
\begin{equation}\label{eq:current}
     I = \frac{G_t}{16e} \int d \epsilon \langle\Tr{\brr{\tau_3 \brr{\check{g}_{\text{L}},\check{g}_{\text{R}}}^\tk}} \rangle_{\vp_F}, 
\end{equation}
from the quasiclassical Green functions, $\check{g}_{\text{L}}$ and $\check{g}_{\text{R}}$, of the left (L) and right (R) superconducting leads. The brackets mean an average over the Fermi surface and the energy integration over $\epsilon$ is performed from $-\infty$ to $\infty$. The structure of the quasiclassical Green functions is detailed below in Sec. \ref{sec:usadel}.

\subsection{Usadel equation for an irradiated SC}\label{sec:usadel}

The quasiclassical Green function (GF) is denoted  $\check{g}=\check{g}(R,t,t')$, where $R$ is the center of mass coordinate and $(t,t')$ are two time arguments. This GF obeys the Usadel equation \cite{Kopnin2001May} 
\begin{equation}\label{eq:usadel}
        -iD\nabla \br{\check{g}\nabla \check{g}} +\{\check{\tau}_3 i\partial_t , \check{g}\} + \Delta(t) e^{i\theta \check{\tau}_3} \brr{i\check{\tau}_2,\check{g}} = 0,
\end{equation} where $D$ is the SC diffusion constant and $\Delta(t)$ the time-dependent gap modulus. The GF has a $4\times 4$ structure in Keldysh-Nambu space and reads
\begin{equation}
\begin{pmatrix}
    \hat{g}^\tr & \hat{g}^\tk \\
    0 & \hat{g}^\ta
\end{pmatrix}    
\end{equation} 
where $\hat{g}^i$ are $2\times 2$ matrices acting in electron-hole Nambu space, with $i=\tr,\ta,\tk$. The upperscript $i=\tr, \ta,\tk$ refers to retarded, advanced and kinetic (or Keldysh) blocks. It is convenient to introduce a set of standard Pauli matrices $\tau_i$ acting in electron-hole Nambu space, and associated enlarged $4\times 4$ matrices, denoted as $\check{\tau}_i = \mathds{1} \otimes \tau_i$, acting in full Keldysh-Nambu space. The superconducting phase is notated $\theta$. The (curly)-bracket corresponds to the (anti)-commutator. The vector potential is introduced into the Usadel equation via the minimal coupling relation 
\begin{equation}
    \nabla \longrightarrow \nabla -ie \brr{\vA(t) \check{\tau}_3, .}.
\end{equation} 
Looking for homogeneous solutions for a bulk superconductor, Eq. (\ref{eq:usadel}) simplifies to 
\begin{align}\label{eq:homogeneous_usadel}
    &iD e^2 \brr{\vA \check{\tau}_3 \circ \check{g} \circ \vA \check{\tau}_3  \circ \check{g} - \check{g} \circ \vA \check{\tau}_3  \circ \check{g} \circ \vA \check{\tau}_3} \nonumber \\ 
    +&\{\check{\tau}_3 i\partial_t , \check{g}\} + \Delta(t) e^{i\theta \check{\tau}_3} \brr{i\check{\tau}_2,\check{g}} = 0,
\end{align} 
with "$\circ$" the convolution operator defined by $(\check{g}\circ  \check{g}) (t,t') = \int dt'' \check{g} (t,t'')\check{g}(t'',t')$ and $\vA (t) \circ \check{g}(t,t') = \vA (t) \check{g}(t,t')$. 
The Usadel equation is completed with the normalization condition 
\begin{equation}\label{eq:norm_condition}
    \check{g}\circ\check{g}(t,t') = \delta (t-t').
\end{equation}
and the SC gap obeys the self-consistent equation 
\begin{equation}
    \Delta(t) = -i\frac{\pi \lambda}{4} \Tr \brr{ \langle \tau_2 \hat{g}^k(t,t) \rangle _{\vp_F}}
\end{equation} with $\langle \dots \rangle _{\vp_F} = \int d\Omega_p / 4\pi$.

In the absence of irradiation, the equilibrium gap is given by the BCS interpolation formula
\begin{equation}
    \Delta_0= \Delta_0 (T) = \Delta_{0,0} \tanh{\brr{1.74 \sqrt{T/T_c - 1}}} \, ,
\end{equation} 
where $T_c$ is the critical temperature and $\Delta_{0,0} \equiv \Delta_0 (T=0)$.

The Higgs mode is a scalar mode that interacts only nonlinearly with the THz light \cite{Shimano2020Mar}. The coupling is second-order with respect to the vector potential and leads to a coherent oscillation of the SC gap at $2\omega$ on top of the static $\Delta_0$
\begin{equation}\label{eq:gap}
    \Delta(t) = \Delta_0 + \Delta_2 e^{-2i\omega t} + \Delta^*_2 e^{2i\omega t},
\end{equation} where $\Delta_2$ is the Higgs mode amplitude. A resonance is obtained for excitation pulsation $\hbar \omega = \Delta_0$ thereby corresponding to a Higgs energy $\hbar \omega_H = 2\Delta_0$ \cite{Varma2002Feb,Pekker2015Mar,Shimano2020Mar}. In our model, the Higgs mode is directly proportional to $\alpha$
\begin{equation}
    |\Delta_2|(\omega) = \alpha f(\omega)
\end{equation} where $f(\omega)$ is a function peaked at the resonance $ \omega = \Delta_0/\hbar$  \cite{Tang2020Jun}.


It is very useful to define the Fourier-Wigner transform of the two-times GF as \cite{Tsuji2008Dec}
\begin{equation}
    \check{g}(t,t') = \sum_{n \in \mathbb{Z}}  \int \frac{d \epsilon}{2\pi} e^{it' \epsilon} e^{-it \epsilon_n} \check{g}_n(\epsilon)
\end{equation} where $\epsilon_n = \epsilon + n \omega$. In the following we refer to $\check{g}_0$ as the time-averaged GF and $\check{g}_n$ as the $n$-th harmonics of the GF. The normalization condition Eq. (\ref{eq:norm_condition}) can be rewritten as 
\begin{equation}
    (\check{g} \circ \check{g})_k (\epsilon) = \sum_{l+l' = k} \check{g}_l(\epsilon_{l'}) \check{g}_{l'}(\epsilon) = \delta_{k,0}.
\end{equation}
We now want to solve order by order the Eq. \eqref{eq:homogeneous_usadel} with respect to the small parameter $\alpha = A_D/\Delta_0 \ll 1$, with $A_D = D e^2 |\vA_0|^2 / \hbar$. We see from Eq. \eqref{eq:homogeneous_usadel} that all $\check{g}_n$ with $n$ odd vanish, as $\vA$ always comes in even powers. In the following, we focus on the second harmonic $\check{g}_{\pm 2}$ which is of order $O(\alpha)$. The higher harmonics $\check{g}_{\pm 2n}$ are smaller and of order $O(\alpha^{n})$. Therefore limiting ourselves to first order in $\alpha$, the two-times GF can be expanded as

\begin{align}
        \check{g}(t,t') &= \int \frac{d \epsilon}{2\pi} e^{it' \epsilon} e^{-it \epsilon} \left[ \check{g}_0(\epsilon) + \check{g}_2(\epsilon) e^{- 2i\omega t} \right. \nonumber  \\ 
     &\left. + \check{g}_{-2}(\epsilon) e^{2i\omega t} \right].
\end{align}

Finally the dc bias $V$ can be included by the following gauge transformation \cite{Larkin1966Nov, Likharev1986Aug, Kopnin2001May}
\begin{equation}\label{subsec:bias}
    \check{g}_V(t,t') =e^{iVt \tau_3}\check{g}(t,t')  e^{-iVt' \tau_3}.
\end{equation}
In the following, we will fix $V_L=0$ and $V_R=V$ as only the potential difference is meaningful. 

\section{Non-irradiated Josephson junction}\label{sec:non_irradiated}

In this section, we present the time-averaged quasi-classical Green functions that serves as building blocks for the results obtained in presence of irradiation (see next section). In the absence of irradiation, the well-known results for the $I(V)$ characteristic of tunnel junctions are recovered.

\subsection{Quasiclassical Green functions $V=0$}

The Usadel equation Eq. (\ref{eq:usadel}) for the retarded $\hat{g}_0^\tr$ component of the GF reads
\begin{align}
    \brr{\epsilon \tau_3 + \Delta_0 i M_{\theta}, \hat{g}_0^\tr} = 0,
\end{align} 
where the Nambu space structure of the gap is encapsulated in the matrix $iM_{\theta} = e^{ i \theta \tau_3} i\tau_2 $.
Using the normalization condition, the zero-order retarded GF is given by \cite{Kopnin2001May}
\begin{equation}\label{eq:g0}
    \hat{g}^\tr_{0,\theta} (\epsilon) = \frac{\epsilon \tau_3 +\Delta_0 i M_{\theta}}{s^{\tr}(\epsilon)},
\end{equation} with $s^{\tr}(\epsilon) = i \sqrt{\Delta_0^2 - (\epsilon + i \gamma)^2}$. Here the parameter $\gamma$, called Dyne parameter \cite{Dynes1978Nov}, is a small phenomenological energy scale that characterizes the depairing effects
in the SC and induces a broadening in the optical response
functions. Mathematically, this parameter is necessary to ensure the proper analytical behavior of the retarded GF.

The advanced GF is related to the retarded by the relation $\hat{g}^\ta = - \tau_3 \br{\hat{g}^\tr}^\dagger \tau_3 $, while the equilibrium Keldysh function at finite temperature $T$ is given by
\begin{equation}
    \hat{g}^\tk_0(\epsilon) = \left( \hat{g}^\tr_0(\epsilon) - \hat{g}^\ta_0(\epsilon) \right) \tanh{\br{\frac{\beta \epsilon}{2}}} \, ,
\end{equation}
where $\beta=1/(k_B T)$ and the phase subscript $\theta$ is implied in the Green functions.

\subsection{Effect of the voltage bias}

We consider that a finite dc bias $V$ is applied between the right SC and left SC. Using gauge invariance, it is possible to consider that the left lead is unaffected while the whole bias $V$ is applied to the right lead : $V_R=V$ and $V_L=0$, while $\theta_R = \chi/2$ and $\theta_L = -\chi/2$. Then Eq. (\ref{subsec:bias}) provides the Fourier-Wigner transform of the retarded GF for the right lead as: 

\begin{equation}
    \hat{g}^\tr_V(\epsilon,\frac{t+t'}{2}) = \frac{\br{\epsilon + V \tau_3}\tau_3}{s^\tr(\epsilon + V \tau_3)} + \frac{e^{i V (t+t') \tau_3} iM_{\chi/2} \Delta_0}{s^\tr(\epsilon)}.
\end{equation}  where a matrix analog of the normalization factor $1/s^{\tr}$ has been defined as
\begin{equation}
    \frac{1}{s^\tr(\epsilon + V \tau_3)} \equiv \begin{pmatrix}
        1/s^\tr(\epsilon + V) & 0 \\
        0 & 1/s^\tr(\epsilon - V)
    \end{pmatrix}.
\end{equation}

In order to compute the current, it is convenient to expand those GF and the normalization matrix in the basis consisting of Nambu space-Pauli matrices as
\begin{align}\label{eq:gv}
    &g_V(\epsilon,t) = g_{V,\mathds{1}}(\epsilon) \mathds{1} + g_{V,3}(\epsilon) \tau_3 
     + e^{2i t V \tau_3} g_{V,\chi}(\epsilon)  iM_{\chi},\\
    &1/s^\tr(\epsilon + V \tau_3) = s_\mathds{1}^\tr(\epsilon,V) ^{-1} \mathds{1} + s_3^\tr(\epsilon,V) ^{-1} \tau_3.
\end{align} 
Finally, one also defines : 
\begin{equation}
\tanh{\br{\beta (\epsilon + V \tau_3)/2}} = f_0 (\epsilon,V) \mathds{1} + f_3 (\epsilon,V)  \tau_3,
\end{equation}
where $f_0$ and $f_3$ are respectively the symmetric and antisymmetric distribution function of the right SC. The Green functions obtained above allow to recover well-known results for the non-irradiated tunnel junction.  

\subsection{Current for the non-irradiated SIS junction}

{\it At zero voltage $V=0$}, the current in a tunnel SIS junction is given by the Ambegaokar-Baratoff formula \cite{Ambegaokar1963Jun} 
\begin{equation}\label{eq:AB63}
    I(\chi) = \frac{G_t}{2e} \pi \Delta_0 \tanh \left(\frac{\beta \Delta_0}{2} \right) \sin \chi
\end{equation} 
where $\Delta_0=\Delta_0(T)$ is the temperature dependent gap. To set the notations for the following calculus, we re-derive this Ambegaokar-Baratoff formula within the quasiclassical formalism.

To obtain the dc current in the junction we need to calculate the commutator 

\begin{align}
    \brr{\check{g}_{0,-\chi/2},\check{g}_{0,\chi/2}}^k = \hat{g}^{\tr}_{0,-\chi/2} \hat{g}^{\tk}_{0,\chi/2} + \hat{g}^{\tk}_{0,-\chi/2} \hat{g}^{\ta}_{0,\chi/2} \nonumber\\
    - \hat{g}^{\tr}_{0,\chi/2} \hat{g}^{\tk}_{0,-\chi/2} - \hat{g}^{\tk}_{0,\chi/2} \hat{g}^{\ta}_{0,-\chi/2}.
\end{align}

From Eq. \eqref{eq:current} we see that only terms proportional to $\tau_3$ will contribute to the current. 

Knowing that

\begin{equation}
    [iM_{-\chi/2},iM_{\chi/2}] = -2i\tau_3 \sin{\chi},
\end{equation} we easily get 

\begin{align}
    \tau_3 \brr{\check{g}_{0,-\chi/2},\check{g}_{0,\chi/2}}^\tk =& -2i\Delta_0^2 \sin{\chi} \br{\frac{1}{(s^\tr)^2}-\frac{1}{(s^\ta)^2}} \nonumber \\ &\times \tanh{\br{\frac{\beta \epsilon}{2}}} + \dots
\end{align}
and after contour integration in Eq. \eqref{eq:current} the dc current is found to be given by Eq. (\ref{eq:AB63}).

{\it At finite bias}, the current through the junction gains ac components at the Josephson pulsation set by $\omega_J = 2eV/\hbar$ and reads: 
\begin{align}
    I(t) &= I_{\text{dc}} +I_0 \sin{\br{\omega_J t + \chi}}+ I'_0 \cos{\br{\omega_Jt + \chi}}\\
        &= I_{\text{dc}} + I_{\text{ac}} \sin{\br{\omega_Jt + \chi + \phi}}
\end{align} with $I_{\text{ac}} = \sqrt{I_0^2 + I^{'2}_0}$ and $\cos{\phi} = I_0 / |I_{\text{ac}}|$. The various current amplitudes depend on the voltage $V$ and are given explicitly by the following expressions: 
\begin{widetext}
\begin{align}
    I_\text{dc}(V) &= \frac{G_t}{8e} \int d \epsilon \ \epsilon \br{\frac{1}{s^\tr} - \frac{1}{s^\ta}} \brr{f_3 \br{g^r_{V,3} - g^a_{V,3}} + \br{g^r_{V,\mathds{1}} - g^a_{V,\mathds{1}}} \br{f_0 - \tanh{\beta \epsilon/2}} }, \label{eq:dc_bias}\\[2ex]
    \begin{split}
    I_0(V) &= -\frac{i\Delta^2_0 G_t}{8e} \int d \epsilon \  \left[ \tanh{\br{\beta \epsilon /2}} \br{\frac{1}{s^\tr} - \frac{1}{s^\ta}} \br{\frac{1}{s_\mathds{1}^\tr(\epsilon,-V)} + \frac{1}{s_\mathds{1}^\ta (\epsilon,-V)}}  \right. \\ &  \left. + f_0(\epsilon,-V) \br{\frac{1}{s^\tr} + \frac{1}{s^\ta}} \br{\frac{1}{s_\mathds{1}^\tr(\epsilon,-V)} - \frac{1}{s_\mathds{1}^\ta (\epsilon,-V)}} +f_3(\epsilon,-V) \br{\frac{1}{s^\tr} + \frac{1}{s^\ta}} \br{\frac{1}{s_3^\tr(\epsilon,-V)} - \frac{1}{s_3^\ta (\epsilon,-V)}} \right],  
\end{split}\\[2ex]
   \begin{split}
         I'_0(V) &=  -\frac{\Delta^2_0 G_t}{8e} \int d \epsilon \  \left[ \tanh{\br{\beta \epsilon /2}} \br{\frac{1}{s^\tr} - \frac{1}{s^\ta}} \br{\frac{1}{s_3^\tr(\epsilon,-V)} - \frac{1}{s_3^\ta (\epsilon,-V)}}  \right. \\ &  \left. - f_0(\epsilon,-V) \br{\frac{1}{s^\tr} - \frac{1}{s^\ta}} \br{\frac{1}{s_3^\tr(\epsilon,-V)} - \frac{1}{s_3^\ta (\epsilon,-V)}} -f_3(\epsilon,-V) \br{\frac{1}{s^\tr} - \frac{1}{s^\ta}} \br{\frac{1}{s_\mathds{1}^\tr(\epsilon,-V)} - \frac{1}{s_\mathds{1}^\ta (\epsilon,-V)}} \right].    
         \end{split} 
\end{align}
\end{widetext}
where $f_0=f_0 (\epsilon,V)$ and $f_3=f_3 (\epsilon,V)$.

\begin{figure}
    \centering
    \includegraphics[width = \columnwidth]{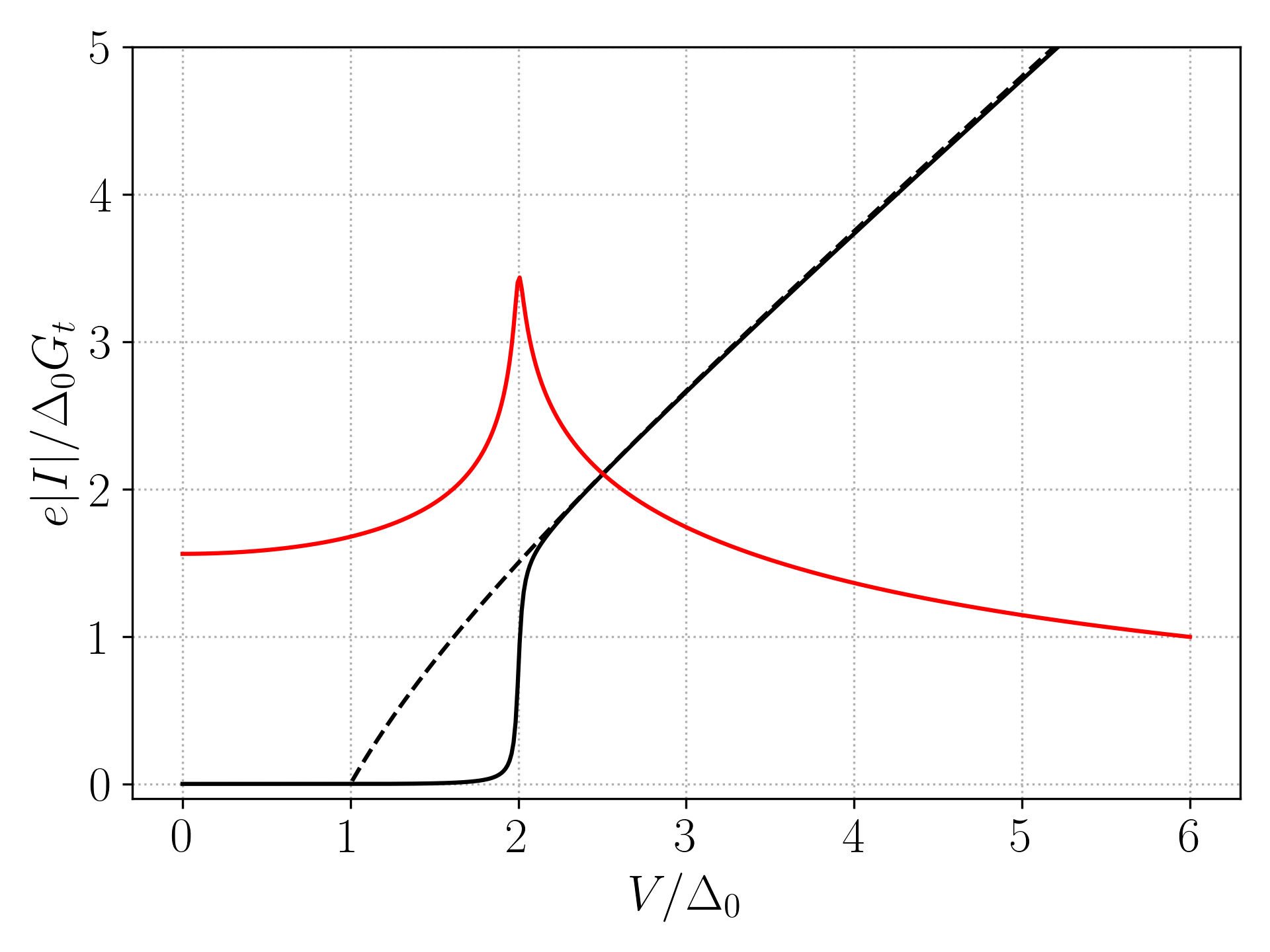}
    \caption{Current $I_\text{dc}$ (solid black curve with a gap) and $I_\text{ac}$ (solid red curve with a peak) as function of the bias $V$. The dashed curve represents the asymptotic limit for the dc current $V (1-\Delta^2/V^2)$ (see Appendix \ref{app:JJ_current}).  Parameters are: $T=0.05T_c$, $\gamma = 0.01 \Delta_{0,0}$.}
    \label{fig:idcac}
\end{figure}

This result has already been derived by Larkin and Ochinnikov \cite{Larkin1966Nov}. The Josephson current amplitude $I_0=I_0(V)$ corresponds to the tunneling of Cooper pairs across the junction, and reduces to Eq. \eqref{eq:AB63} at $V=0$ as expected. The current $I_{\text{dc}}$ is independent of $\chi$ and comes from the tunneling of quasiparticles across the junction, while $I'_0$ is the phase-dependent part of the quasiparticle current \cite{Josephson1962Jul, Langenberg1974Jan, Harris1974Jul}. The terms proportional to $I_0$ and $I'_0$ are recasted into the ac component with amplitude  $I_{\text{ac}}$.
As expected, for $eV=0$ only $I_0$ is non-zero. We cannot evaluate those integrals analytically except in certain regimes (see Appendix \ref{app:JJ_current} for more detail). 
The dc current is almost zero as long as $eV < 2\Delta_0$ (Fig. \ref{fig:idcac}), as the potential is not strong enough to actually break a Cooper pair. The ac current presents a peak at $eV=2\Delta_0$, known as the Riedel peak \cite{Hamilton1971Feb, Harris1974Jul}, that becomes a logarithmic singularity at $eV=2\Delta_0$ in the limit $\gamma \rightarrow 0$. It originates from the density of state singularity at the gap. 

\section{Irradiated Josephson junction }\label{sec:irradiated} 

In this section, we present our results describing the effect of irradiation on the junction in regimes where the Higgs mode can be excited. 

\subsection{Second-harmonics Green function}
We first compute the second-harmonics $g_2$ for the Green function in the irradiated lead, which requires a first order calculation in irradiation strength $\alpha$.
To first order in $\alpha$, the Usadel equation Eq. (\ref{eq:homogeneous_usadel}) reads 
\begin{align}\label{eq:g0alpha}
    \brr{\epsilon \tau_3 + \Delta_0 i M_{\theta}, \hat{g}_0^\tr} + i \Delta_0 \alpha \left[\Bar{\hat{g}}^\tr_0 (\epsilon_{-1}) + \Bar{\hat{g}}^\tr_0 (\epsilon_{1}), \hat{g}_0^\tr \right] = 0.
\end{align} with $\Bar{\mathcal{O}} = \tau_3 \mathcal{O} \tau_3$.
In Appendix \ref{app:delta_correction}, it is shown that the $\alpha$-correction to Eq. \eqref{eq:g0} is very weak. Moreover, in the perturbative expression of the second harmonics $\check{g}_{\pm 2}$, only the GF $\check{g}_{0}$ at $\alpha=0$ is consistent to stay at first order in $\alpha$. We can therefore neglect the $\alpha$-correction in \eqref{eq:g0} and so the time-averaged GF stays the same as in Eq. (\ref{eq:g0}) (more details in App. \ref{app:delta_correction}).

In order to disentangle the physics directly to the irradiation from the Higgs mode effects, we decompose the second harmonics of the retarded GF as 
\begin{equation}
\hat{g}^\tr_2(\epsilon) = \hat{g}^\tr_{2,A}(\epsilon) + \hat{g}^\tr_{2,\Delta}(\epsilon) \, ,
\end{equation}
where $\hat{g}^\tr_{2,A}(\epsilon)$ is directly proportional to the field strength and independent of the amplitude of the Higgs mode $\Delta_2$, while $\hat{g}^\tr_{2,\Delta}(\epsilon)$ is the contribution due exclusively to the presence of the Higgs mode.  
We find \cite{Tang2020Jun}

    \begin{align}
        \hat{g}^\tr_{2,A}(\epsilon) &=  \frac{i\Delta_0 \alpha}{s^\tr(\epsilon_2) + s^\tr(\epsilon) } \brr{\Bar{\hat{g}}^\tr_0(\epsilon_1) -\hat{g}^\tr_0(\epsilon_2) \Bar{\hat{g}}^\tr_0(\epsilon_1) \hat{g}^\tr_0(\epsilon)},\\
    \hat{g}_{2,\Delta}^\tr(\epsilon) &= \frac{\Delta_2}{s^\tr(\epsilon_2) + s^\tr(\epsilon)} \brr{iM_{-\chi/2} - \hat{g}^\tr_0(\epsilon_2) iM_{-\chi/2} \hat{g}^\tr_0(\epsilon)}.
\end{align} 
Finally those $2 \times 2$ matrices are expanded in Nambu space Pauli matrices as  
\begin{equation}
\hat{g}_{2,i}(\epsilon) = \tau_3 g_{23,i}(\epsilon)  + i\tau_2 g_{22,i}(\epsilon) \, .
\end{equation}
Furthermore upon the change of variables $\epsilon \rightarrow \epsilon - V\tau_3$, the function $g_{22,i}(\epsilon)$ becomes the following $2 \times 2$ matrix: 
\begin{equation}
g_{22,i}(\epsilon - V\tau_3)  =  g_{22,\mathds{1},i}(\epsilon,V) \mathds{1} + g_{22,3,i}(\epsilon,V) \tau_3
\end{equation}
with $i = A, \Delta$. These functions $g_{22,\mathds{1},i}(\epsilon,V)$ and $g_{22,3,i}(\epsilon,V)$ will appear on the current formula below
(Eq. \eqref{eq:i23}).
\bigskip

\subsection{Tunneling current}

In presence of an electromagnetic driving at pulsation $\omega$ and finite bias $V$, the total tunneling current gain further harmonics resulting from the interplay of the Josephson pulsation $\omega_J$ and the drive pulsation $\omega$:  
\begin{align}\label{eq:i(t)}
    I(t) &= I_{\text{dc}} + I_{\text{ac}} \sin{\br{\omega_J t + \chi + \phi}} \nonumber \\ &+
    2 \sum_{i=A,\Delta} |I'_{2,i}| \cos{\br{2\omega t - \theta'_i}} \cos{\br{\omega_J t + \chi}}  \nonumber \\
    &+
    2 \sum_{i=A,\Delta} |I_{2,i}| \cos{\br{2\omega t - \theta_i}} \sin{\br{\omega_J t + \chi}}  \nonumber \\
    & +2\sum_{i=A,\Delta} |I_{3,i}| \cos{\br{2\omega t - \theta_{3,i}}}.
\end{align}
where we have introduced the index $i=A,\Delta$ to distinguish contributions proportional to $\alpha$ or $\Delta_2$. To better characterize the spectral content of the time-dependent current Eq. \eqref{eq:i(t)} it is useful to linearize the  cosines/sinus products. The term proportionnal to $ \cos{\br{2\omega t - \theta'_i}} \cos{\br{\omega_J t + \chi}}$  and the one proportional to $\cos{\br{2\omega t - \theta'_i}} \sin{\br{\omega_J t + \chi}} $ both yield pulsations $\omega_I = \pm (2\omega \pm \omega_J)$. 

Hence, the spectrum of the ac current exhibits the following pulsations : $\omega_I = \pm (2\omega \pm \omega_J), \pm 2\omega$. In particular, this leads to a dc component contribution when $2\omega = \pm \omega_J$ or $eV = \pm \hbar \omega$ coming from the second order current in $|A_0|$, whose amplitude is denoted $I_2$. This results in Shapiro peaks \cite{Shapiro1964Jan} in the $I_{dc}(V)$ current.
The various current amplitudes depend both on voltage $V$ and pulsation $\omega$ (Fig. \ref{fig:i2}), and are given by the following explicit formula:

\begin{widetext}
    \begin{align}\label{eq:i23}
\begin{split}
    I_{2,i}(V) &= -\frac{i\Delta_0 G_t}{8e} \int d \epsilon \ \left[ \tanh{\br{\beta \epsilon /2}} \br{\frac{1}{\Omega^r} - \frac{1}{\Omega^a}} g^r_{22,\mathds{1},i} + \tanh{\br{\beta \epsilon_2 /2}} \br{\frac{1}{\Omega^r(\epsilon_2)} - \frac{1}{\Omega^a(\epsilon_2)}} g^a_{22,\mathds{1},i}  \right. \\ &  \left. + \br{\frac{1}{\Omega^r(\epsilon_2)} + \frac{1}{\Omega^a}} \br{f_0(\epsilon,-V)  g^r_{22,\mathds{1},i} + f_3(\epsilon,-V)  g^r_{22,3,i} - f_0(\epsilon_2,-V)  g^a_{22,\mathds{1},i} - f_3(\epsilon_2,-V)  g^a_{22,3,i} + g^{\text{an}}_{22,\mathds{1},i}} \right]    \end{split} \\[2ex]
       \begin{split}
         I'_{2,i}(V) &=  -\frac{\Delta_0 G_t}{8e} \int d \epsilon \ \left[ \tanh{\br{\beta \epsilon /2}} \br{\frac{1}{\Omega^r} - \frac{1}{\Omega^a}} g^r_{22,3,i} - \tanh{\br{\beta \epsilon_2 /2}} \br{\frac{1}{\Omega^r(\epsilon_2)} - \frac{1}{\Omega^a(\epsilon_2)}} g^a_{22,3,i}  \right. \\ &  \left. - \br{\frac{1}{\Omega^r(\epsilon_2)} - \frac{1}{\Omega^a}} \br{f_0(\epsilon,-V)  g^r_{22,3,i} + f_3(\epsilon,-V)  g^r_{22,\mathds{1},i} - f_0(\epsilon_2,-V)  g^a_{22,3,i} - f_3(\epsilon_2,-V)  g^a_{22,\mathds{1},i} + g^{\text{an}}_{22,3,i}} \right]    \end{split} \\[2ex]
    I_{3,i}(V) &= -\frac{G_t}{8e} \int d \epsilon g^k_{2,3,i} \br{g^r_{V,\mathds{1},i} - g^a_{V,\mathds{1},i}} + g^k_{V,\mathds{1},i}(\epsilon_2) g^a_{2,3,i} - g^r_{2,3,i} g^k_{V,\mathds{1},i}.
\end{align} 
\end{widetext}

The expression of the anomalous GF $\hat{g}_2^\text{an}$ can be found in Appendix \ref{app:usadel}. The $\theta^{(')}_i$ phase shift is defined with the relation $\cos{\theta^{(')}_i} = \mathfrak{Re}(I^{(')}_{2,i})/|I^{(')}_{2,i}|$. 
In this system the ac current caused by the irradiation consists of two contributions: one explicitly proportional to $\alpha$ and directly caused by the irradiation ($I_{A}$) and another induced by the Higgs mode and proportional to $|\Delta_2|$ ($I_{\Delta}$).






\subsection{Resonance and spectroscopy}
\begin{figure}
    \centering
    \includegraphics[width=\columnwidth]{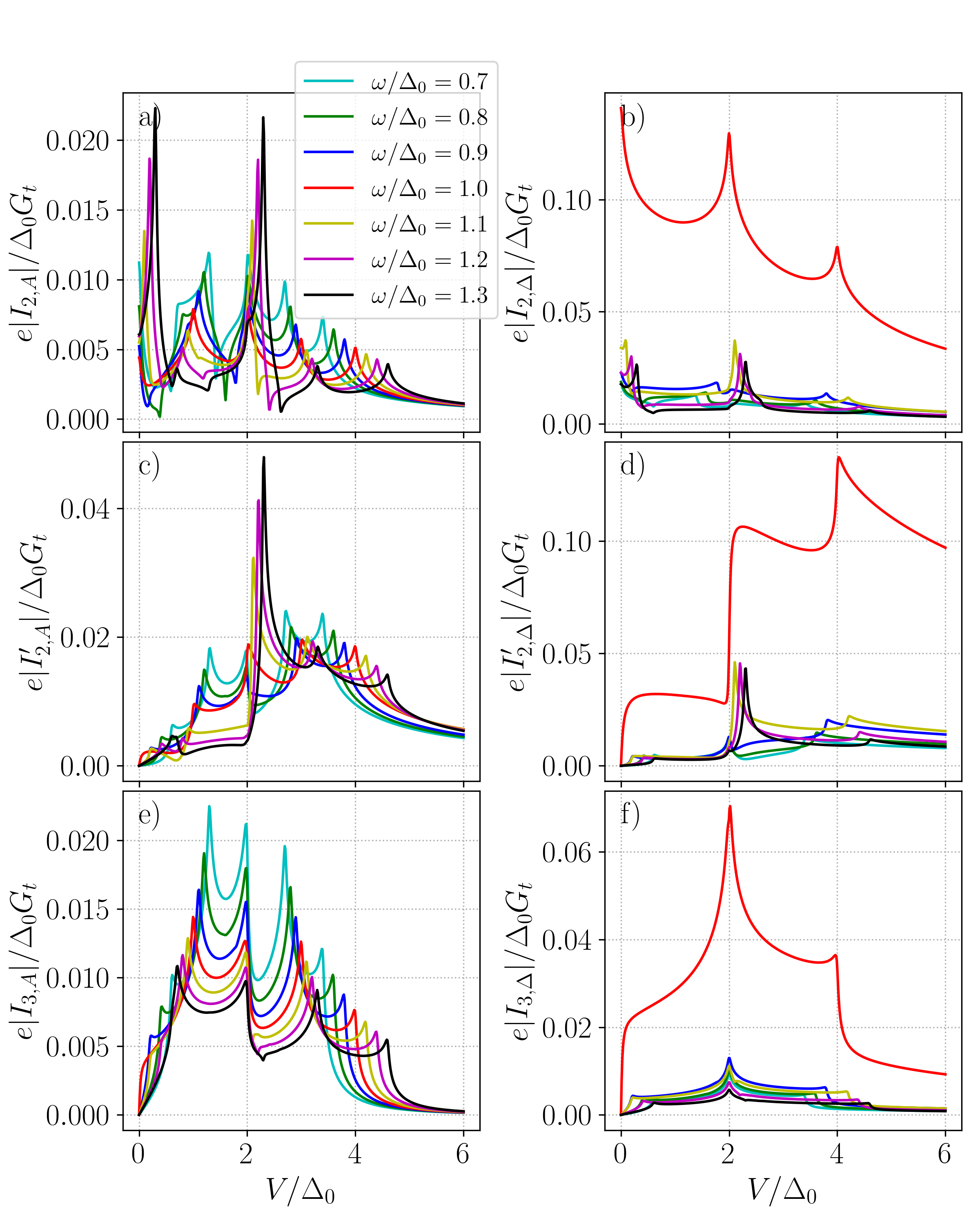}
    \caption{Current amplitudes given by \eqref{eq:i(t)} as function of the bias $V$ for different irradiation frequencies $\omega$. {\it Left panel}: Currents directly proportional to the irradiation parameter $\alpha$. {\it Right panel}: Currents due to the Higgs mode, which are clearly resonant at $\hbar \omega=\Delta_0$. Different resonances peaks appear at specific voltages $V$ due to PAT. Note the different scales on the vertical current axes. Parameters are: $T=0.05T_c$, $\gamma = 0.01 \Delta_{0,0}$, $\alpha = 0.01$.}
    \label{fig:i2}
\end{figure}

The amplitudes of the different contributions to the current are plotted in Fig. \ref{fig:i2} as function of the voltage $V$ for various pulsations $\omega$. As a main result, at the resonant condition $\hbar \omega = \Delta_0$, the current generated by the Higgs is always way larger than the other contributions. This is due to the fact that $\Delta_2$ is resonant at this frequency \cite{Tang2020Jun}.

Moreover, we further predict a "fine structure" of this resonant current as function of bias voltage. For the currents $I_{2,A}$ and $I'_{2,A}$ we expect to get peaks at bias $eV= 2\Delta_0 + m \hbar \omega$ due to PAT \cite{Cuevas2002Mar,Haxell2023Oct}, with $m\in \mathbb{Z}$. In Fig. \ref{fig:i2} we indeed observe peaks at bias $eV = 2\Delta_0 - 2\hbar\omega,\ 2\Delta_0 - \hbar \omega,\ 2\Delta_0, \ 2\Delta_0 + \hbar \omega, 2\Delta_0 + 2\hbar \omega$ for $\hbar \omega \leq \Delta_0$. In the case of $I_{2,A}$ we have additional peaks at $eV = \hbar \omega,\ 2\hbar \omega$. This can be understood as PAT of Cooper pairs tunneling. When $\hbar \omega > \Delta_0$, we have additional peak at $eV = \Delta_0 + \hbar \omega$ (and also $eV = \hbar \omega - \Delta_0$ for $I_{2,A}$). Those large peaks signal resonances in the QPs tunneling currents, due to the BCS-DOS singularity shifted by the PAT. 

The current $I_{3,\Delta}$ can be compared to the Higgs current calculated in NIS system in \cite{Tang2020Jun}. If we set $\Delta_0 = 0$ in the right SC, the $I_{3,\Delta}$ reduces to the Higgs current obtained in \cite{Tang2020Jun}. In our case a peak appears at $eV = 2\Delta_0$ which is the mass of the amplitude mode (Fig. \ref{fig:i2}). Since the Higgs mode can be interpreted as the pairing and un-pairing of electrons and holes around the gap at frequency $2\Delta_0$, the current will be maximum if the bands of the 2 superconductors are biased by a dc potential $eV=2\Delta_0$, thus explaining this fixed pic at all frequencies. At $eV = 4\hbar \omega$ a kink appears due to sidebands effect. Contrary to the other currents contributions, the peak at $eV=2\Delta_0$ is always the maximum current for any $\omega$, sign that this current comes from only the Higgs mode.

It is interesting to compare this peaks $I(V)$ structure of the current amplitudes due to the Higgs (Fig. \ref{fig:i2} Right Panel) with respect to peaks appearing in the current due to A (Fig. \ref{fig:i2} Left Panel). Looking at the Josephson currents $I^{(')}_{2,\Delta}$ in Fig. \ref{fig:i2}, one notices that half the peaks are missing compared to the $A$-currents. The resonant bias are now at $eV = 2\Delta_0 \pm 2\hbar\omega$ for $\hbar \omega \leq \Delta_0$ plus $eV=2\hbar\omega$ in the case of $I_{2,\Delta}$.  This is coherent with the fact that the Higgs mode can only interact with light to second order such that it can only participate to half the photons assisted transport compared to the $A$-current. This helps to disentangled the contribution due to the mere irradiation of the SC and the contribution specifically due to the generation of the Higgs mode.

Note that the effects plotted in Fig. \ref{fig:i2} are obtained for the particular value of the dimensionless irradiation strength $\alpha=0.01$, and are proportional to the light intensity $I$. For $\alpha=0.01$, $\Delta_0 = 1 $ meV, $D=10^{-3}$ m$^2$.s$^{-1}$ \footnote{A diffusion constant $D \sim 10^{-3}$ m$^2$.s$^{-1}$ is obtained for a mean free path $l \sim 1 nm$ and Fermi velocity $v_\text{F} = 10^{6}$m.s$^{-1}$. Those orders of magnitude are expected for really dirty SCs \cite{Clarke1968Feb}.}, $\omega = 1.5$ THz (which corresponds to $1$ meV), one obtains an intensity I $\sim 1$ W.cm$^{-2}$. This corresponds to an electrical field between $1$ and $10$ kV.m$^{-1}$ while typical fields are around 300 kV.m$^{-1}$ in \cite{matsunaga_2014_NbN}.  
Note that the observation of these effects should not be affected by the presence of the Josephson plasma resonance which depends on the capacitance of the junction and arises at GHz frequencies \cite{Dahm1968Apr, Siddiqi2005Jan}. 

\subsection{Particular case : $V=0$}
Let us discuss the particular case of the unbiased irradiated JJ. Even for $V=0$ a finite $I_{2,\Delta}$ amplitude is predicted (Fig. \ref{fig:i2}). This is in contrast with previous result in NIS system \cite{Tang2020Jun} where a bias was necessary to produce Higgs mode current. This new current term can be understood as an ac correction to the Josephson current due to the modulation of one of the gap modulus. In fact, the current in a JJ of 2 SCs with two different but closed constant gaps is given by $I_0 \propto \pi \Delta_0 \Delta'_0 /2 (\Delta_0+ \Delta'_0)$. If we take one of the gap to be time dependant such as $\Delta'_0 = \Delta_0 + 2\Delta_2 \cos{2\omega t}$ we except to get a current $\propto \pi \Delta_0/2 + \pi \Delta_2 \cos{\br{2\omega t}}$ which is what we observe. Indeed at $V=0$ the current \eqref{eq:i(t)} reduces to 

\begin{equation}
    I(t) = \brr{I_0 + \sum_{i=A,\Delta} |I_{2,i}| \cos{\br{2\omega t - \theta_i}}}\sin{\chi}.
\end{equation} Near resonance, $|I_{2,\Delta}| \gg |I_{2,A}|$ and numerically we find $e|I_{2,\Delta}|/G_t \Delta_0 \sim \pi |\Delta_2|/\Delta_0$, as expected.

\section{Conclusion}

We have investigated the interplay of Higgs mode physics and Josephson effect in tunneling SIS junctions between conventional superconductors subjected to an electromagnetic irradiation. Using the Keldysh-Usadel formalism, we have evaluated the contribution to the current which is directly generated by the Higgs mode. Various signatures of the Higgs mode excitation are found in the $I(V)$ characteristics. These currents are very sensitive to the irradiation frequency and reach a maximum at the resonant condition $\hbar\omega = \Delta_0(T)$, corresponding to excitation of the Higgs mode. This resonance can be obtained by tuning the frequency or the temperature. Even at zero bias, we predict a non-zero ac current due to irradiation and whose magnitude is mostly due to the Higgs mode excitation at resonance $\omega=\Delta_0(T)/\hbar$. Moreover a Shapiro peak proportional to the Higgs mode develops at bias $eV = \hbar \omega$ and should be directly measurable in the dc current. Additional resonant current peaks, originating from photon-assisted-transport, appear at biases $eV = 2\Delta_0(T) + m \hbar \omega$ in the $I(V)$ curves, with $m$ integer. Being selectively enhanced by the Higgs mode resonance, the even-$m$ peaks are expected to be much stronger than the odd-$m$ peaks.

\section{Aknowledgements}
We thank Alexander Buzdin and Vadim Plastovets for very useful discussions.
This work was supported by the ANR SUPERFAST and the “LIGHT S\&T Graduate Program” (PIA3 Investment for the Future Program, ANR-17-EURE-0027) and GPR LIGHT.

\bibliographystyle{apsrev4-2}
\bibliography{biblio.bib}

\appendix

\section{\texorpdfstring{Irradiation $\alpha$}{TEXT}-correction for \texorpdfstring{$\hat{g}^{\tr/\ta}_0$}{TEXT}}\label{app:delta_correction}

The Usadel equation for the time-averaged GF is 

\begin{align}\label{eq:usa}
    \brr{\epsilon \tau_3 + \Delta_0 i M_{\pm \chi/2}, \hat{g}_0^\tr} + i \Delta_0 \alpha \left[\Bar{\hat{g}}^\tr_0 (\epsilon_{-1}) + \Bar{\hat{g}}^\tr_0 (\epsilon_{1}), \hat{g}_0^\tr \right] = 0.
\end{align}

Let us first look at the simpler situation where we impose an dc current bias, i.e. $\omega = 0$. The equation slightly simplifies into

\begin{align}\label{eq:usadel_0freq}
    \brr{\epsilon \tau_3 + \Delta_0 i M_{\pm \chi/2}, \hat{g}_0^\tr} + i \Delta_0 \alpha \left[\Bar{\hat{g}}^\tr_0 , \hat{g}_0^\tr \right] = 0
\end{align} where we redefined $2\alpha \rightarrow  \alpha$.

In the case $\chi = 0$ we can always write 

\begin{equation}
    \hat{g}_0^\tr(\epsilon) = G_0(\epsilon) \tau_3 + F_0(\epsilon) i\tau_2, 
\end{equation} with $G_0$ and $F_0$ some functions. The normalization condition can be written as $G_0 ^2 - F_0^2 = 1$. We can rewrite Eq. \eqref{eq:usadel_0freq} as an algebraic equation for, e.g., the function $F_0$

\begin{equation}
    -4\Delta_0^2\alpha^2 F_0^4 + 4i \alpha \Delta_0^2(F_0^3+F_0) + F_0^2 (\Delta_0^2 - 4 \alpha^2 - \epsilon^2)  + \Delta_0^2 = 0.
\end{equation} We discard the second order terms in $\alpha^2$ and using the approximation $F_0^3+F_0 \simeq F_0^3 \epsilon^2/\Delta_0^2 $ we get 

\begin{equation}
    4i \alpha \epsilon^2  F_0^3 -(\epsilon^2-\Delta_0^2)  F_0^2  + \Delta_0^2 = 0. 
\end{equation} When $|\epsilon/\Delta_0| \ll 1$ we can use a perturbative approximation in $\alpha$ to find roots of the equation. Defining $F_{\alpha} (\epsilon) = F_{0,\alpha}  + \alpha F_{1,\alpha} $, we find 

\begin{align}
    F_{0,\alpha}  &= \frac{\Delta_0}{s^\tr(\epsilon)} \\ F_{1,\alpha}  &= -2i\frac{\epsilon^2 \Delta_0^2}{s^\tr(\epsilon)^4}.
\end{align}

\begin{figure}
    \centering
    \includegraphics[width=\columnwidth]{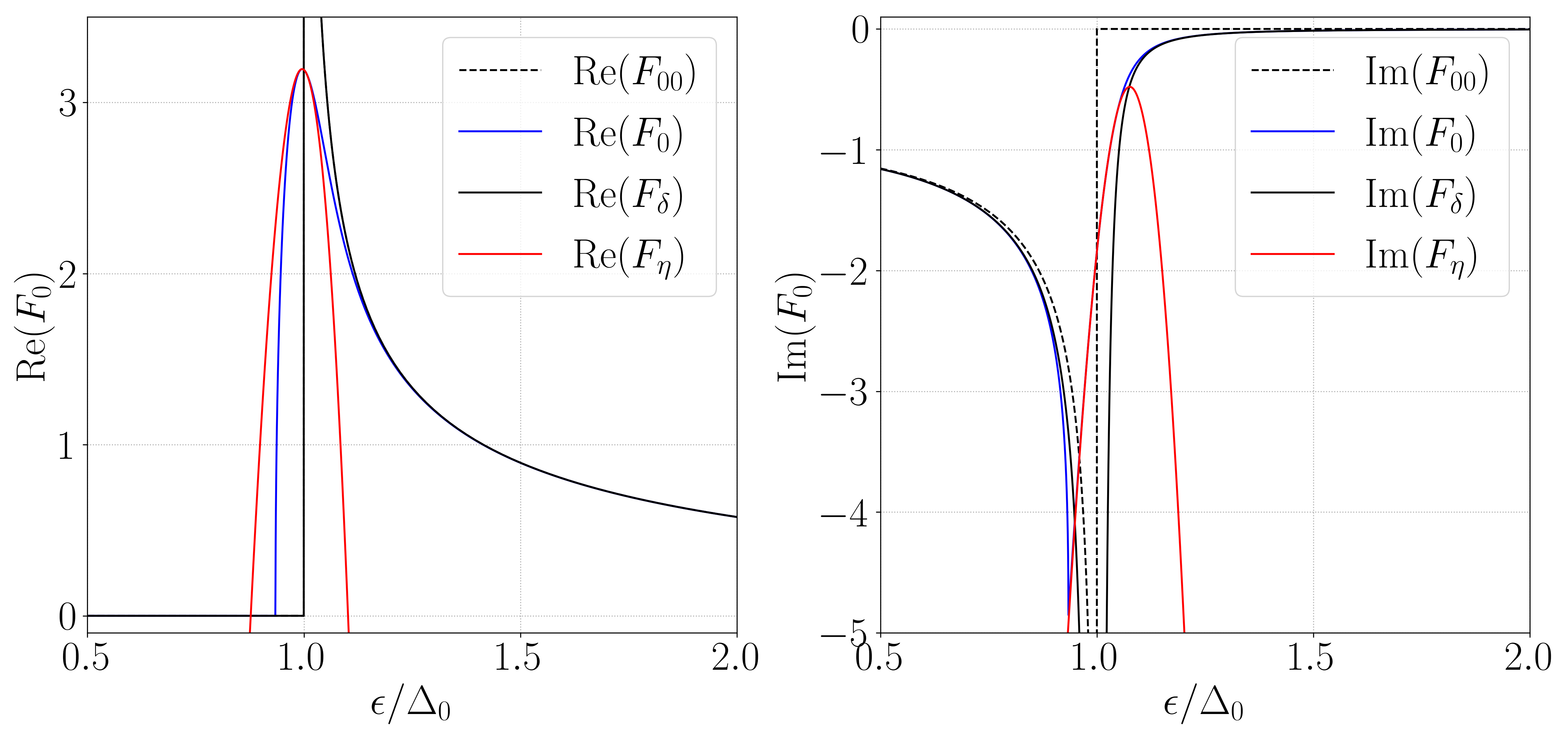}
    \caption{Real/Imaginary part of the Green function $F_0(\epsilon)$ for $2\alpha=0.01$, $\omega = 0$. The function $F_{00}$ is defined as $F_0(\alpha=0)$. The blue curve corresponds to the exact numerical calculation. The red (black) curves are approximations correct at low $\eta$ ($\alpha$).}
    \label{fig:f0}
\end{figure}

Around $|\epsilon/\Delta_0| \approx 1$, we need to use $\eta = (\epsilon^2-\Delta_0^2)/\Delta_0^2$ as perturbative parameter. Defining $F_{\eta} (\epsilon) = F_{0,\eta}+ \eta F_{1,\eta} + \eta^2 F_{2,\eta} $, we find 

\begin{align}
    F_{0,\eta} &= e^{-i\pi/6}\br{\frac{\Delta_0}{4\alpha}}^{\frac{1}{3}} \\ 
    F_{1,\eta} &= \frac{i\Delta_0}{12 \alpha} - e^{-i\pi/6} \frac{1}{3} \br{\frac{\Delta_0}{4\alpha}}^{\frac{1}{3}}\\
    F_{2,\eta} &= \frac{-\Delta_0^{5/3} 4^{1/3}}{144} \left[\frac{e^{i\pi/6}}{\alpha^{5/3}} + 3i  \br{\frac{4}{\Delta_0}}^{2/3} \frac{1}{\alpha} \right.\nonumber \\ &\left. -2 \brr{\frac{e^{-i\pi/6}}{\alpha^{1/3}}}\br{\frac{4}{\Delta_0}}^{4/3}\right]. 
\end{align} We see that those solutions are non-perturbative in $\alpha$, as expected. We plotted in Fig. \ref{fig:f0} the GF $F_0$ and its approximations. When the Dynes parameter $\gamma \gg \alpha \Delta_0$ the first order approximation in $\alpha$ becomes a good approximation, even for $\epsilon  \sim \Delta_0$.

\begin{figure}
    \centering
    \includegraphics[width=\columnwidth]{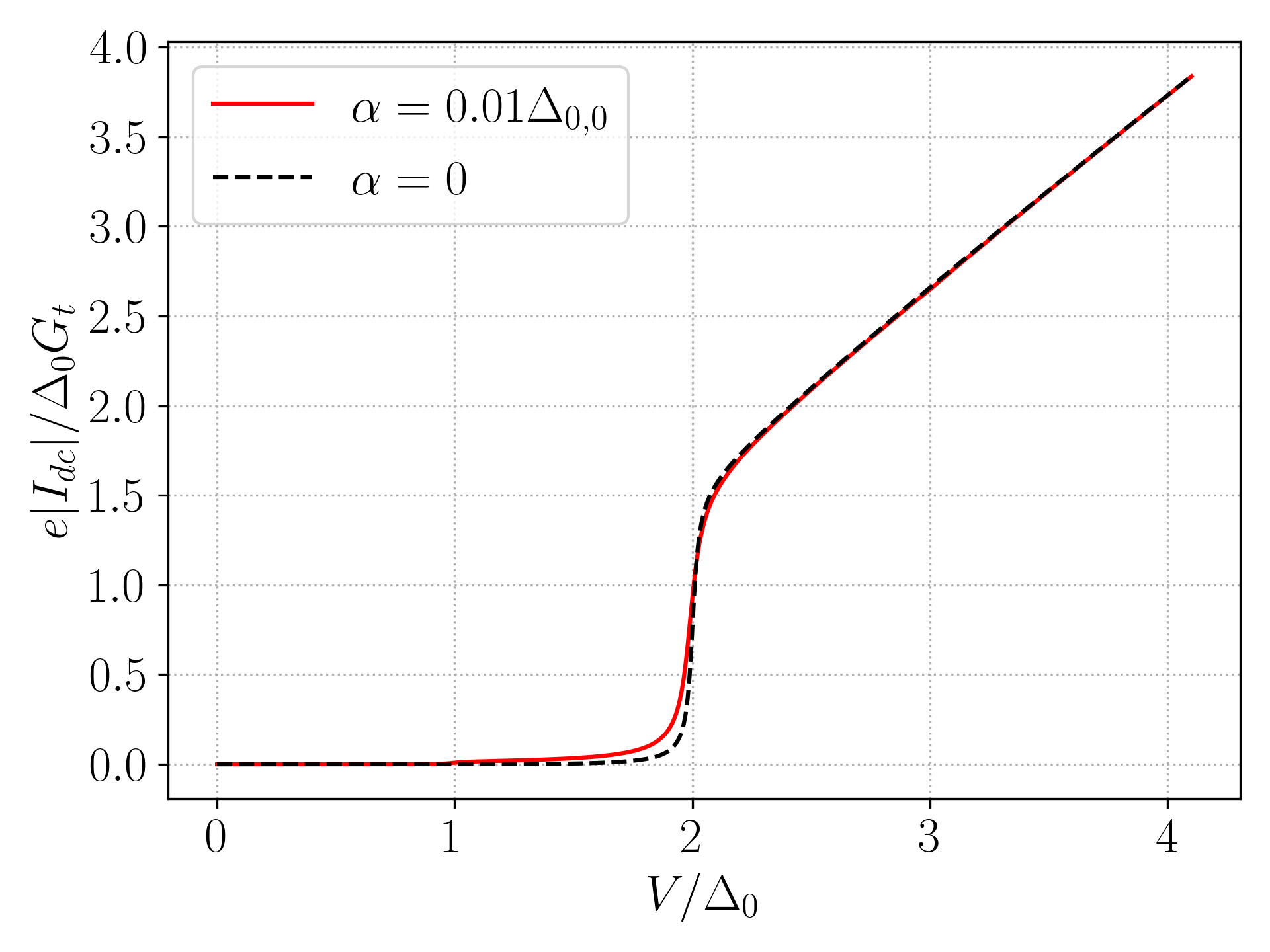}
    \caption{Current $I_\text{dc}$ with and without irradiation. The correction is relatively small for $\alpha = 0.01$. Here $\gamma = 0.01 \Delta_{0,0}0$ and $\omega = \Delta_0$}
    \label{fig:ialpha}
\end{figure}

Analytical results have been obtain in the case $\omega >0$ in the regime $\gamma \gg \alpha \Delta_0$ \cite{Semenov2020Feb}.  
We find in this case 

\begin{equation}
    i F_{1,\alpha} = \frac{\Delta_0^2 \epsilon}{s^\tr(\epsilon_1) s^\tr(\epsilon)^3}\br{\epsilon_1 + \epsilon} + \frac{\Delta_0^2 \epsilon}{s^\tr(\epsilon_{-1}) s^\tr(\epsilon)^3}\br{\epsilon_{-1} + \epsilon}.
\end{equation}
In the case $\gamma \sim \alpha \Delta_0$ we need to solve numerically Eq. \eqref{eq:usa}. We used the algorithm given in \cite{Semenov2016Jul}. The dc current correction $I_\text{dc}(\alpha)$ is given in Fig. \ref{fig:ialpha}. We get it by replacing $\epsilon/s^\tr(\epsilon)- \epsilon/s^\ta(\epsilon)$ in Eq. \eqref{eq:dc_bias} by $2 \mathfrak{Re}(G^\tr_0(\epsilon))$. Knowing that the density of states (DOS) $N(\epsilon) = N_0 \mathfrak{Re}(G^\tr_0(\epsilon))$ \cite{Kopnin2001May,Semenov2016Jul}, with $N_0$ the normal DOS, we see that the $\alpha$-correction comes from a modulation of the DOS. We recover, at least qualitatively, the results of \cite{Cuevas2002Mar} in the limit of low transmission. The correction remains small and our approximation is justified. 

\section{Keldysh Green function}\label{app:usadel}

To find the Keldysh function it is usefull to write the solution as a sum of regular and anomalous term 
\begin{equation}
    \hat{g}^\tk_i (\epsilon) = \hat{g}^{\text{reg}}_i(\epsilon) + \hat{g}^{\text{an}}_i(\epsilon) 
\end{equation} with $g^{\text{reg}}_i(\epsilon) = \hat{g}^\tr_i (\epsilon) h_0(\epsilon) - h_0(\epsilon_i) \hat{g}^\ta_i (\epsilon)$ where the distribution function $h_0 (\epsilon) = \tanh{\br{\beta \epsilon / 2}}$. Solving the Usadel equation for $\hat{g}^\text{an}_2$ we find 

\begin{align}
        \hat{g}^\text{an}_{2,A}(\epsilon) &=  \frac{i\Delta_0 \alpha \brr{\Bar{\hat{g}}^\tr_0(\epsilon_1) -\hat{g}^\tr_0(\epsilon_2) \Bar{\hat{g}}^\ta_0(\epsilon_1) \hat{g}^\ta_0(\epsilon)}}{s^\tr(\epsilon_2) + s^\ta(\epsilon) } \nonumber \\
        &\times \brr{\tanh{\beta \epsilon_2 /2} - \tanh{\beta \epsilon_1 /2}} \nonumber \\
        &+\frac{i\Delta_0 \alpha \brr{\Bar{\hat{g}}^\tr_0(\epsilon_1) -\hat{g}^\tr_0(\epsilon_2) \Bar{\hat{g}}^\tr_0(\epsilon_1) \hat{g}^\ta_0(\epsilon)}}{s^\tr(\epsilon_2) + s^\ta(\epsilon) } \nonumber \\
        & \times \brr{\tanh{\beta \epsilon_1 /2} - \tanh{\beta \epsilon /2}} \\
    \hat{g}_{2,\Delta}^\tr(\epsilon) &= \frac{\Delta_2}{s^\tr(\epsilon_2) + s^\ta(\epsilon)} \brr{iM_{-\chi/2} - \hat{g}^\tr_0(\epsilon_2) iM_{-\chi/2} \hat{g}^\ta_0(\epsilon)} \nonumber \\
    &\times \brr{\tanh{\beta \epsilon_2 /2} - \tanh{\beta \epsilon /2}}.
\end{align}

\section{Low temperature limits of the Josephson current}\label{app:JJ_current}

At $T=0$ and for $V < 2 \Delta_0$, we can rewrite $I_0$ as 
\begin{align}
    \frac{8e I_0}{G_t \Delta_0^2} &= 8 \int^{V+\Delta_0}_{\Delta_0} d\epsilon \frac{1}{\sqrt{\epsilon^2 - \Delta_0^2}} \frac{1}{\sqrt{\Delta_0^2 - (\epsilon-V)^2}} \\
    &= \frac{8}{V} \int^{1}_{0} d\epsilon \frac{1}{\sqrt{(\epsilon + \Delta_0/V)^2 - (\Delta_0/V)^2}} \nonumber\\
    &\times \frac{1}{\sqrt{(\Delta_0/V)^2 - (\epsilon-1 + \Delta_0/V)^2}} \\
    &= \frac{8}{V} \int^{1}_{0} d\epsilon \frac{1}{\sqrt{-\epsilon (\epsilon + 2 \Delta_0/V)}} \nonumber \\
    &\times \frac{1}{\sqrt{(\epsilon - 1 + 2 \Delta_0/V) (\epsilon -1)}}.
\end{align}

To solve this we do the change of variable 

\begin{equation}
    \Bar{\epsilon} = \sqrt{\frac{2 \Delta_0/V \epsilon}{\epsilon -1 + 2 \Delta_0/V}}
\end{equation}

knowing that 

\begin{equation}
    d \Bar{\epsilon} = d \epsilon \frac{2 \Delta_0 / V (2 \Delta_0 / V - 1)}{2 (\epsilon -1 + 2 \Delta_0/V)^{3/2} \sqrt{2 \Delta_0/V \epsilon}}.
\end{equation} We get

\begin{align}
    \frac{8e I_0}{G_t \Delta_0^2} &= \frac{8\sqrt{2\Delta_0/V} }{\Delta_0 (2 \Delta_0 / V - 1)} \int^{1}_{0} d\Bar{\epsilon}\frac{\epsilon -1 + 2 \Delta_0/V}{\sqrt{ (\epsilon + 2 \Delta_0/V)(1-\epsilon)}} \\
    &= \frac{8}{\Delta_0} \int^{1}_{0} d\Bar{\epsilon}\frac{1}{\sqrt{ 1-\Bar{\epsilon}^2}}\frac{1}{\sqrt{ 1-\frac{V^2}{4\Delta_0^2}\Bar{\epsilon}^2}}\\
    &= \frac{8}{\Delta_0} \text{K}\br{\frac{V^2}{4\Delta_0^2}},
\end{align}
with $\text{K}(x)$ the complete elliptic integral of the first kind defined as 

\begin{equation}\label{eq:defK}
    \text{K}(x) = \int_0^1 dt \frac{1}{\sqrt{1-t^2}} \frac{1}{\sqrt{1-x t^2}}.
\end{equation}

When $V>2\Delta_0$ we have

\begin{align}
    \frac{8e I_0}{G_t \Delta_0^2} &= 8 \int^{V+\Delta_0}_{V-\Delta_0} d\epsilon \frac{1}{\sqrt{\epsilon^2 - \Delta_0^2}} \frac{1}{\sqrt{\Delta_0^2 - (\epsilon-V)^2}} \\
    &= \frac{16}{V}\text{K}(\frac{4\Delta_0^2}{V^2}).
\end{align}

From Eq. \eqref{eq:defK} we see that $\text{K}(0) = \frac{\pi}{2}$. We find the correct limit when $V=0$. In the large bias limit $V \gg \Delta_0$ we get 

\begin{equation}
    I_0 = \frac{G_t}{e}\pi \frac{\Delta^2_0}{V}.
\end{equation}

\begin{figure}
    \centering
    \includegraphics[width=\columnwidth]{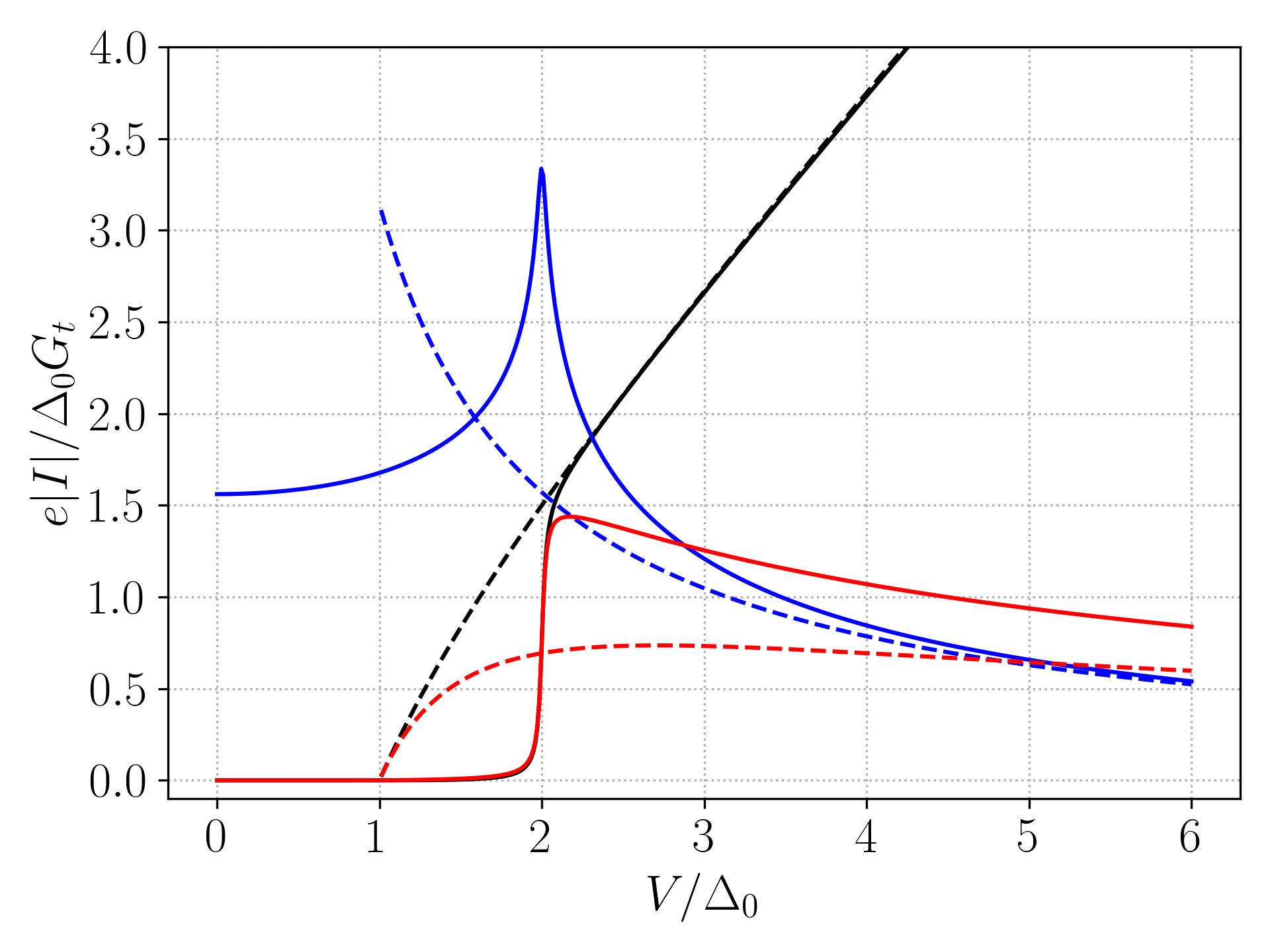}
    \caption{Currents $I_\text{dc}$ (solid black curve), $I_0$ (solid blue curve) and $I'_0$ (solid red curve) as a function of the dc bias $V$. The dashed plots correspond to the asymptotic limits valid for $V\gg 2\Delta_0$. Parameters : $T=0.05T_c$, $\gamma = 0.01 \Delta_{0,0}$.}
    \label{fig:iv0}
\end{figure}

Now let us look at the low temperature limit for $I_\text{V}$ and $I'_0$. We rewrite them as  
\small
\begin{align}
      \frac{8e I_\text{V}}{G_t} &= -2\br{\int^{V-\Delta_0}_{\Delta_0} +  2\int^{\infty}_{V+\Delta_0}}d\epsilon \frac{|\epsilon|}{\sqrt{\epsilon^2 - \Delta_0^2}} \nonumber \\
      &\times \frac{|\epsilon-V|}{\sqrt{(\epsilon-V)^2 - \Delta_0^2}} \brr{\tanh{\br{\beta (\epsilon-V)/2}} - \tanh{\br{\beta \epsilon/2}}}\\
    \frac{8e I'_0}{G_t \Delta_0^2} &= -2\br{\int^{V-\Delta_0}_{\Delta_0} +  2\int^{\infty}_{V+\Delta_0}} d\epsilon \frac{\sign{\epsilon}}{\sqrt{\epsilon^2 - \Delta_0^2}} \nonumber\\
    &\times \frac{\sign{(\epsilon-V})}{\sqrt{(\epsilon-V)^2 - \Delta_0^2}} \brr{\tanh{\br{\beta (\epsilon-V)/2}} - \tanh{\br{\beta \epsilon/2}}}.
\end{align}
\normalsize
When $V\gg \Delta_0,\ T$ we get (Fig. \ref{fig:iv0})

\begin{align}
      \frac{8e I_\text{V}}{G_t} &\simeq 4 \int^{V-\Delta_0}_{\Delta_0} d\epsilon \frac{|\epsilon|}{\sqrt{\epsilon^2 - \Delta_0^2}} \frac{|\epsilon-V|}{\sqrt{(\epsilon-V)^2 - \Delta_0^2}}  \\
      & = 8 V \int^{1/2}_{\Delta_0/V} dx \frac{x}{\sqrt{x^2 - \frac{\Delta_0^2}{V^2}}} \frac{1-x}{\sqrt{(1-x)^2 - \frac{\Delta_0^2}{V^2}}}\\
      &\sim 8V \br{1 - \frac{\Delta_0^2}{V^2}}, \\
    \frac{8e I'_0}{G_t \Delta_0^2} &\simeq -4\int^{V-\Delta_0}_{\Delta_0} d\epsilon \frac{1}{\sqrt{\epsilon^2 - \Delta_0^2}} \frac{1}{\sqrt{(\epsilon-V)^2 - \Delta_0^2}} \\
     & = -\frac{8}{V}\int^{1/2}_{\Delta_0/V} dx \frac{1}{\sqrt{x^2 - \frac{\Delta_0^2}{V^2}}} \frac{1}{\sqrt{(1-x)^2 - \frac{\Delta_0^2}{V^2}}}\\
    &\sim -\frac{16}{V} \ln{\br{V/\Delta_0}}.
\end{align}

\end{document}